\documentclass[aps,10pt]{revtex4}
\usepackage{epsfig,graphics,amssymb,amsmath,subeqnarray,amsthm,latexsym}

\def\u{{\bf u}}\def\bka{{\boldsymbol \kappa}}\def\tr{{\rm tr}}\def\d{{\rm d}}\def\p{\partial}\def\btau{\boldsymbol{\tau}} \def\ex{{\bf e}_x}\def\ey{{\bf e}_y}
\def\bgam{\boldsymbol{\gamma}}\def\gamd{\dot\bgam}\def\I{{\bf 1}}\def\W{{\cal W}}
\def\dgamd{\stackrel{\triangledown}{\gamd}}\def\dbtau{\stackrel{\triangledown}{\btau}} \def\dsgamd{\stackrel{\Box}{\gamd}}\def\dsbtau{\stackrel{\Box}{\btau}}\def\De{{\rm De}}\def\tp{\widetilde\psi} 

\begin{document}
\title{Propulsion in a viscoelastic fluid}
\author{Eric Lauga\footnote{Email: lauga@mit.edu}}
\affiliation{Department of Mathematics,
Massachusetts Institute of Technology,
77 Massachusetts Avenue, 
Cambridge, MA 02139.}
\date{\today}
\begin{abstract}

Flagella beating in complex fluids are significantly influenced by viscoelastic stresses. Relevant examples include the ciliary transport of respiratory airway mucus and the motion of spermatozoa in the  mucus-filled female reproductive tract. We consider the simplest model of such propulsion and transport in a complex fluid, a waving sheet of small amplitude free to move in a polymeric fluid with a single relaxation time. We show that, compared to self-propulsion in a Newtonian fluid occurring at a velocity $U_N$, the sheet swims (or transports fluid) with velocity 
$$
\frac{U}{U_N} = \frac{1+\De^2{\eta_s}/{\eta}}{1+\De^2} ,
$$
where   $\eta_s$ is the viscosity of the Newtonian solvent, $\eta$ is the zero-shear-rate viscosity of the polymeric fluid, and $\De$ is the Deborah number for the wave motion, product of the wave frequency by the fluid relaxation time. Similar expressions are derived for the rate of work of the sheet and the mechanical efficiency of the motion. These results are shown to be independent of the particular nonlinear constitutive equations chosen for the fluid, and are valid for both waves of tangential and normal motion. The generalization to more than one relaxation time is also provided. In stark contrast with the Newtonian case, these calculations  suggest that transport and locomotion in a non-Newtonian fluid can be conveniently tuned without having to modify the waving gait of the sheet but instead by passively modulating the material properties of the liquid.


\end{abstract} 
\maketitle

\section{Introduction}\label{intro}

G.I. Taylor all but started a new field half a century ago when he showed that a two-dimensional waving sheet could self-propel in a viscous fluid in the absence of inertia  \cite{taylor51}.  Much progress toward understanding small-scale biological locomotion has been made  since Taylor's contribution, and a number of classical reviews address  the fundamental fluid mechanical issues associated with swimming microorganisms \cite{lighthill76,brennen77,childress81,braybook}. In this paper, motivated by the observation that a variety of internal biological fluids have complex polymeric microstructures and display non-Newtonian behavior, we revisit  Taylor's original calculation in the case where the fluid is non-Newtonian.

We start by reviewing  two examples where complex fluids are involved in biological locomotion and transport, namely ciliary transport of respiratory mucus and sperm motion in the mucus-filled female reproductive tract. After summarizing the modeling approaches used in the past to quantify the influence of viscoelasticity on these examples, we motivate and introduce the model used in this paper.

\subsection{Biological background}
\subsubsection{Ciliary transport of respiratory mucus}
The first example of a viscoelastic  biological fluid is the mucus present in our upper respiratory tract. It is a glycoprotein-based 
gel \cite{bansil95,lillehoj02}  which is transported along the epithelium by large arrays of beating cilia, allowing removal of foreign substances and a healthy respiratory system \cite{sleigh88}. Cilia are short flagella which produce fluid motion by means of a collaborative beating process (metachronal  waves) and  have long been recognized as crucial in many biological transport processes, in particular for respiratory and reproductive functions, as well as sensing mechanisms and propulsion \cite{gibbons81}. 

In the upper respiratory tract, cilia are located in a thin layer of  low-viscosity Newtonian fluid (serous layer) and penetrate partially in the overlying high-viscosity non-Newtonian mucous blanket. The rheology of respiratory airway mucus is critical to ensure normal transport conditions and many respiratory diseases such as asthma or cystic fibrosis lead to rheological changes  \cite{hwang69}. Early rheological measurements of respiratory mucus reported a large shear viscosity,  on the order of 1000 Pas, and a
 relaxation time scale  $\lambda\approx$ $100-1000$ s \cite{denton68,hwang69}. Later measurements showed a larger range of viscosities, between 10 and 1000 Pas \cite{gilboa76,shake87}, with significantly shorter relaxation times, $\lambda\sim 30 $ s  \cite{gilboa76}, strong shear-thining behavior \cite{shake87} and elastic modulus on the order of $1-10$ Pa \cite{giordano78, meyer80,wolf80}. Given that common cilia oscillate at frequencies of about $f  = \omega/2\pi \approx  $ 5-50 Hz \cite{brennen77}, the fluid mechanics of ciliated transport in respiratory mucus occurs therefore in the limit where the Deborah number, $\De= \lambda \omega$, is much larger than one, and  elastic effects are therefore predominant.

\subsubsection{Motion of spermatozoa in the female reproductive tract}

The second example of non-Newtonian biological fluid arises when we consider the locomotion of spermatozoa in the female reproductive tract. During their journey to the ovum, sperm cells are both actively swimming and passively transported by peristaltic-like flows (inside the uterus). Along the way, they encounter a variety of viscoelastic fluids, through the cervix (glycoprotein-based  cervical mucus), inside the uterine cavity, at the uterotubal junction, and inside the fallopian tubes (mucosal epithelium) where cells are stored and undergo chemical and mechanical changes \cite{suarez06}. The  cumulus oophorus, thin matrix  covering the outside of the ovum, is also a viscoelastic (actin-based) gel \cite{dunn76}. These complex fluids act as mechanical barriers for the selection of healthy spermatozoa and, together,  lead to a decrease of six orders of magnitude in the size of the population of swimming cells, from tens of millions reaching the cervix to only tens approaching the ovum.

Early rheological studies  showed  that the cervical mucus viscosity is about $100$ Pas, with  relaxation times on the order of $\lambda\approx 100$ s \cite{denton68,hwang69}. Subsequent measurements reported smaller relaxation times, between 1 and 10 s \cite{eliezer74,litt76,wolf77_2,tam80}, elastic modulus on the order of $0.1-10$ Pa \cite{eliezer74,litt76,meyer76, wolf77_1,wolf77_2} and viscosities in the range $ 0.1  - 10$ Pas \cite{litt76,meyer76, wolf77_1,wolf77_2}. These number vary during the  menstrual cycle \cite{wolf77_3} and depend strongly on the hydration of the mucus \cite{tam80}. Given that the flagella of spermatozoa typically beat at a frequency between 20 and 50 Hz  \cite{brennen77}, spermatozoa swimming in mucus occurs therefore also at large Deborah numbers. The rheology of cumulus oophorus was characterized as well, with relaxation time of less than half a second for small deformation \cite{dunn76}.  The microstructure of cervical mucus  was later examined by microscopy \cite{yudin89}, and revealed a crosslinked gel with a typical length scale between polymeric filaments below one micron.

The viscoelastic forces resulting from a complex mucus rheology significantly influence the waveforms, structure and swimming path of spermatozoa \cite{fauci06}.  Early experimental study of the motion of human sperm cells in cervical mucus showed that, compared to motion in semen,  their flagella beat with higher frequency, smaller amplitude and smaller wavelength, resulting in unchanged swimming velocity but  straighter paths \cite{katz78}. A subsequent study of the motion of bull spermatozoa in cervical mucus also reported lower swimming speed per unit beat than in a viscous fluid, as well as important wall effects \cite{katz81}.  
 
A last important issue regarding the motion of spermatozoa in complex fluids is called ``hyperactivation''. It is a change in their flagellar beat pattern  which occurs during their stay on the mucus-filled epithelium of the Fallopian tubes. The beat becomes asymmetric and results in higher force generation,  helping the cells detach from the mucosal epithelium, enhancing their mobility in viscoelastic media, and facilitating penetration of the matrices surrounding the oocyte  \cite{suarez91,suarez96,ho01,suarez03}.  Experimental study of the motion of active and hyperactive spermatozoa in polymer solutions showed that hyperactive sperm swim faster than active (normal) cells as the elasticity of the solution increases (the viscosity remaining constant) and that hyperactive sperms swim in straight lines only in viscoelastic media \cite{suarez92}.

\subsection{Modeling approaches in flagellar and ciliary fluid mechanics}

We now turn to the modeling of locomotion and transport in biological fluids. Much work has been devoted to the fluid mechanics of flagella and cilia in Newtonian flows, and the reader is referred to classical texts for reviews \cite{blake74,lighthill76,brennen77,childress81}. Quantitative approaches for ciliary flow include infinite models, where the envelope of the cilia tips is replaced by an infinite waving two-dimensional or three-dimension surface \cite{taylor51,blake71b}, envelope models  where an impenetrable oscillating surface is embedded on a finite three-dimensional object \cite{blake71a,brennen74}, and  sublayer models which detail the fluid forces at the individual flagella level \cite{blake72,gueron92,gueron93,gueron97,gueron99}. Classical modeling approaches in single-flagella hydrodynamics include resistive-force theory, where the fluid forces are described by a local anisotropic drag tensor \cite{gray55,lighthill76,brennen77} and slender-body hydrodynamics \cite{tillett70,batchelor70,cox70,keller76-jfm,geer76,lighthill76,johnson80} where  long and thin filaments are represented by a line distribution of flow singularities (see also Ref.~\cite{fauci06} for an introduction to recent numerical work).

A number of modeling approaches for mucus transport and sperm locomotion in complex fluids have been proposed in the past. Respiratory mucus transport was quantified by two-fluid models, all showing that the mechanical properties of the mucous layer are critical for efficient transport \cite{ross74,blake75,blake80,yates80,liron83}. Taylor's original calculation of propulsion by a waving sheet \cite{taylor51,tuck68} was adapted to a variety of  configurations relevant to a complex fluids, in particular in the case of interaction  with oscillating walls \cite{smelser74}, interaction  with  a phenomenologically active material \cite{shukla78}, motion in a confined environment \cite{katz80} and in porous media \cite{siddiqui03,siddiqui05}  (see also related work on presitaltic transport of complex fluids \cite{siddiqui91,siddiqui93,siddiqui94,hayat04}). Numerical computations for the motion of flagellated cells interacting with deformable walls were reported  \cite{fauci95}, showing  swimming velocities decreasing with an increase in the wall elasticity.

Only a few studies incorporated the viscoelastic properties of the fluid in a continuum-mechanics approach. The first effect of fluid elasticity was quantified using a retarded-motion expansion and assuming a second-order fluid \cite{chaudhury79}. A more rigorous integral constitutive relationship was also used \cite{sturges81}. For zero Reynolds number, the swimming velocities reduce to the Newtonian velocities in both studies \cite{chaudhury79, sturges81}.  A phenomenological modification to the equations of the resistive-force theory of slender-body hydrodynamics was proposed as a model for flagella motion in a linear viscoelastic fluid \cite{fulford98}. In that limit, the swimming velocity is unchanged from the Newtonian value, but the work done by the swimmer always decreases, and a complex fluid is seen to increase the swimming efficiency. A somewhat similar model was recently used to model changes in flagella waveforms in beating elastic filaments \cite{fu06}.

\subsection{Purpose of the study and structure of the paper}

In this paper, we propose to take a rigorous modeling approach to the fluid rheology in order to obtain quantitative information on transport and locomotion in a viscoelastic fluid in the biologically-relevant  limit of large Deborah numbers. We consider Taylor's waving sheet as a idealized geometrical model of both spermatozoa locomotion (free sheet, quiescent flow in the far field) and ciliary transport (fixed sheet, finite induced velocity in the far field) in complex fluids. Although this is clearly a simplified model, it allows us to treat rigorously the effect of viscoelastic stresses on transport processes, and is  therefore likely to  shed some light on the biological examples detailed above.

Using two models derived from the kinetic theory of polymers, we demonstrate that the viscoelastic nature of the fluid modifies in a non-trivial manner  the kinematics and energetics of swimming of the sheet and transport of the surrounding fluid. Compared to the case of a purely Newtonian solvent, swimming and transport with the same gait always require more work and lead to lower speeds. Our results are therefore in stark contrast with those of Ref.~\cite{fulford98} where a linear viscoelastic material was considered.  This arises because, generically, the net magnitude of locomotion and transport scale quadratically with the amplitude of the local oscillatory motion, and therefore nonlinear terms in non-Newtonian constitutive relationships can never be neglected. Surprisingly, we also show that our results are unchanged if we consider other classical empirical constitutive relationships to describe the fluid dynamics. One implication of our results is that, although non-Newtonian fluids lead to a deterioration of the transport performance,  they allow on the other hand biological systems to tune passively transport kinematics by modulating material properties. This feature is notably absent in Newtonian fluids and might suggest an advantageous biological control mechanism.
 
The structure of the paper is the following. In \S\ref{setup} we present the setup and notation for the swimming sheet model. The swimming velocity of the free sheet (or induced far-field velocity for a fixed sheet) and its rate of work  are then calculated in  \S\ref{oldroyd} and \S\ref{FENE} for the Oldroyd-B and FENE-P models with a single relaxation time respectively. A discussion of the results and their relevance to biology is presented in \S\ref{discussion}. In Appendix \ref{retarded}, we re-derive the results for  a second-order fluid. We then show in Appendix \ref{JSO} and \ref{G} that other classical rheological models  (Johnson-Segalman-Oldroyd  and  Giesekus) lead to the same results as those obtained in the main section of this paper. Furthermore, we  show in Appendix \ref{tangential} that our results are unchanged if we consider a sheet with traveling waves of both normal and  tangential motion. Finally, the main results of this paper are generalized in Appendix \ref{manymodes} to the case where the viscoelastic fluid possesses more than one relaxation time.


\section{Setup}\label{setup}

\begin{figure}[t]
\centering
\includegraphics[width=0.6\textwidth]{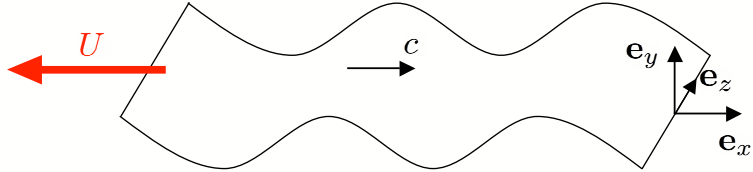}
\caption{Schematic representation and notation for the swimming sheet.}
\label{mainfig}
\end{figure}

We consider in this paper the simplest model for an oscillatory swimmer, analogous to that proposed by  Taylor in his original paper \cite{taylor51}. An infinite inextensible sheet of negligible thickness is swimming steadily in an incompressible viscoelastic fluid (see Fig.~\ref{mainfig} for notations). In the frame moving at the (yet unknown) swimming speed, the vertical position of the sheet, described by the function $y(x,t)$, oscillates in time according $y(x,t)=a \sin (k x - \omega t)$. This profile is a traveling wave of amplitude $a$, wavelength $\lambda = 2\pi/k$ and period $T=2\pi/\omega$ along the positive $x$ direction (w.l.o.g  both $k$ and $\omega$ are assumed to be positive) and is responsible for the swimming motion. If the sheet is not free to move, a velocity opposite to the swimming velocity will be induced in the far field, and therefore the same calculation provides models for both locomotion and transport in a complex fluid.

We nondimensionalize lengths by $1/k$, times  by $\omega$, and velocities by the traveling wave speed, $c=\omega/k$, so that the dimensionless vertical position of the sheet is written $y(x,t)=\epsilon \sin (x - t)$, where we have defined $\epsilon = ak$. In this paper, we will assume that the amplitude of the sheet is much smaller than the wavelength and will derive the swimming results  in the asymptotic limit $\epsilon \ll 1$  \footnote{The inextensibility condition leads to tangential motion of order $\epsilon^2$ along the sheet and has therefore no influence on the leading order swimming kinematics \cite{childress81}.}.

\subsection{Kinematics}


Since the sheet is two dimensional, we will look to solve the problem using a streamfunction, $\psi(x,y,t)$, so that the two components of the velocity field, ${\bf u}$,  are given by $u_x=\p \psi / \p y$, $u_y=-\p \psi / \p x$, and the incompressibility condition, $\nabla \cdot {\bf u} = 0$, is always satisfied. The boundary conditions for the streamfunction arise from conditions at infinity and  on the waving sheet.  Let us denote by $-U {\bf e}_x$ the steady swimming speed of the sheet, where we use a minus sign because we expect swimming to occur in the direction opposite to that of the traveling wave as in the Newtonian case. In the frame moving at the steady swimming speed, the velocity field far from the sheet asymptotes therefore to $U {\bf e}_x$, leading to the boundary conditions
\begin{equation}
\nabla \psi \big\vert_{(x,\infty)}  = U\ey.
\end{equation}
On the swimming sheet, the no-slip boundary condition leads to 
\begin{equation}\label{bc_sheet}
\nabla \psi\big\vert_{(x,\epsilon \sin(x-t))}   = \epsilon \cos (x-t) \ex.
\end{equation}
We look for a regular perturbation expansion in powers of $\epsilon$ for both the streamfunction and the swimming 
velocity, in the form
\begin{equation}
\psi = \epsilon \psi_1 +  \epsilon^2 \psi_2 + ...,\quad U=\epsilon U_1 + \epsilon^2 U_2 + ..., 
\end{equation}
and because of the symmetry $\epsilon \to -\epsilon$, we expect swimming to happen at best at order $\epsilon^2$. 
Expanding the boundary conditions on the sheet, Eq.~\eqref{bc_sheet}, we find that the leading order problem 
satisfies
\begin{subeqnarray}\label{bc_order1}
\nabla \psi_1 \big\vert_{(x,\infty)}  & = & U_1 \ey,  \\
\nabla \psi_1 \big\vert_{(x,0)} & = &   \cos (x-t)\ex.
\end{subeqnarray}
while the boundary conditions for the stream function at order $\epsilon^2$ are
\begin{subeqnarray}\label{bc_order2}
\nabla \psi_2 \big\vert_{(x,\infty)}  & = &  U_2\ey,\\
\nabla \psi_2 \big\vert_{(x,0)} & = & -\sin(x-t) \nabla \left(\frac{\p \psi_1}{\p y}\right) \bigg\vert_{(x,0)}.
\end{subeqnarray}

\subsection{Kinetics}
We assume the Reynolds number for the fluid motion to be very small, and write therefore  mechanical equilibrium of the fluid as
\begin{equation}\label{mecheq}
\nabla p = \nabla \cdot \btau
\end{equation}
where $p$ is the pressure and $\btau$ the (deviatoric) viscoelastic stress tensor. The pressure can be eliminated from the mechanical equilibrium by taking the curl of Eq.~\eqref{mecheq} and considering therefore the resulting vorticity equation, which is a fourth-order partial differential equation for $\psi$, with boundary conditions given by Eqs.~\eqref{bc_order1} and \eqref{bc_order2} for the first two orders in $\epsilon$. 

In order to close the system of equations and solve the problem we need to have a constitutive equation relating stresses to kinematics. Following classical monographs \cite{doi88,bird76,birdvol1,birdvol2,larson88,tanner88,bird95,larson99}, we decompose the viscoelastic stress, $\btau$, into two  components, $\btau = \btau^s+\btau^p$, where $\btau^s$ is the Newtonian contribution from the solvent,  $\btau^s=\eta_s \gamd$, with $\gamd = \nabla\u + \nabla\u ^T $, and  $\btau^p$ is the polymeric stress contribution.  Modeling choices govern the evolution equation for $\btau^p$.

\section{Moderate Deborah numbers: Oldroyd-B}
\label{oldroyd}

As our first model, we consider non-Newtonian fluids governed by the Oldroyd-B constitutive equation with a single relaxation time \cite{oldroyd1950,bird76,birdvol1,birdvol2,larson88,tanner88,bird95,larson99}. This is arguably the most famous constitutive equation for polymeric fluids, and for two reasons. Firstly, it is very simple, and as so is very attractive. Secondly, it can be exactly derived from the kinetic theory of polymers  in the case where the polymer molecules are modeled by elastic dumbbells, and therefore is physically sound. Oldroyd-B fluids agree  reasonably well with experimental measurements of  polymeric flows up to order one Weisenberg numbers (${\rm We} = \lambda \dot{\gamma}\sim 1$), but is known to possess a number of unphysical features at large values of extension rates when elastic dumbbells become infinitely elongated \cite{oldroyd1950,bird76,birdvol1,birdvol2,larson88,tanner88,bird95,larson99}.

\subsection{Model}

In the Oldroyd-B model, the polymeric stress, $\btau^p$ satisfies the upper-convected Maxwell equation
\begin{equation}\label{UCM}
\btau^p + \lambda  {\stackrel{\triangledown}{\btau^p}} = \eta_p \gamd,
\end{equation}
where $\lambda$ is the polymer relaxation time  and $\eta_p$ the polymer contribution to the viscosity. The relaxation time measures the typical rate of decay of stress when the fluid is at rest and is zero for a Newtonian fluid. The upper-convected derivative, defined for the tensor ${\bf A}$ as 
\begin{equation}
\stackrel{\triangledown}{{\bf A}} = \frac{\p {\bf A}}{\p t} + {\bf u}\cdot \nabla{\bf A} - (\nabla \u^T\cdot {\bf A} + {\bf A}\cdot \nabla \u),
\end{equation}
is the rate of change of ${\bf A}$ calculated while translating and deforming with the fluid.
It is then straightforward to see that Eq.~\eqref{UCM} can be manipulated  to obtain the Oldroyd-B constitutive equation for the total stress, $\btau$, as
\begin{equation}
\btau +\lambda_1 \dbtau= \eta [\gamd + \lambda_2 \dgamd ],
\end{equation}
where $\eta=\eta_s+\eta_p$, $\lambda_1=\lambda$, and $\lambda_2=\eta_s\lambda/\eta<\lambda_1$, and which 
is the form we will use to solve the swimming problem. Note that  $\lambda_2$ is termed the retardation time scale and is equal to the rate of decay of residual rate of strain when the fluid is stress-free.
 
We now nondimensionalize stresses by $\eta \omega$ and shear rates by $\omega$, leading to the constitutive relationship 
\begin{equation}\label{oldroydB}
\btau +\De_1 \dbtau= \gamd + \De_2 \dgamd,
\end{equation}
where we have defined the two Deborah numbers $\De_1=\lambda_1 \omega$ and $\De_2=\lambda_2 \omega$ and where we use the same symbols for convenience.

\subsection{Order $\epsilon$ solution and rate of work}
\label{oldroyd_order1}

We look for a solution for the stress as a regular perturbation expansion in powers of $\epsilon$,
\begin{equation}
\btau = \epsilon \btau_1 + \epsilon^2 \btau_2+...,
\end{equation}
and the  leading order solution of Eq.~\eqref{oldroydB} is written as
\begin{equation}\label{oldroydB_order1}
\btau_1 +\De_1 \frac{\p  \btau_1}{\p t}= \gamd_1 + \De_2 \frac{\p \gamd_1}{\p t}\cdot
\end{equation}
Since Eq.~\eqref{oldroydB_order1} is a linear equation, we can first calculate its divergence and then its curl to obtain the equation for the leading order vorticity, and we get
\begin{equation}
\left(1+ \De_2 \frac{\p }{\p t}\right) \nabla^4 \psi_1 = 0.
\end{equation}
Given the boundary conditions from Eq.~\eqref{bc_order1}, and using Fourier transforms, it  is straightforward to see that  $\psi_1$ is the same as for the Newtonian case,  that is
\begin{equation}\label{taylor_1}
\psi_1(x,y,t)=(1+y)e^{-y}\sin (x-t),\quad U_1=0.
\end{equation}
At leading order in $\epsilon$, the flow field and swimming kinematics are therefore unchanged by the presence of viscoelastic stresses. The rate of work of the sheet, however, is modified. Let us use Fourier notations, and write 
\begin{equation}\label{order1_sol}
\psi_1=\Re\left\{\tp_1e ^{it} \right\}, \quad \tp_1=i(1+y)e^{-y}e^{-ix} .
\end{equation}
Using similar notations for the stress and rate-of-strain tensors, we obtain the first-order stress as given by
\begin{equation}\label{stress_1}
\widetilde \btau_1 = \frac{1+i\De_2}{1+i\De_1} \widetilde \gamd_1 \cdot
\end{equation}
The leading-order rate of viscous dissipation, written as ${\cal W} = \epsilon^2 {\cal W}_2 + ...$,
is equal to the volume integral of $\btau_1:\gamd_1$ on both sides of the sheet, which average value in Fourier notation is the volume integral of 
\begin{equation}\label{mean_work}
\langle \btau_1:\gamd_1 \rangle = \frac{1}{2}\Re\{ \widetilde\btau_1: \widetilde\gamd_1^*\}=
\frac{1}{2}\left(\frac{1+\De_1\De_2}{1+\De_1^2}\right)[ \widetilde\gamd_1: \widetilde\gamd_1^*] ,
\end{equation}
where $\langle... \rangle$ denotes the average over one wavelength of the sheet. As a consequence, the leading order value of the rate of viscous dissipation, equal to the rate of work done by the sheet, 
is given by, in dimensional variables and per unit length in the $z$ direction (see Fig.~\ref{mainfig}),
\begin{equation}\label{work_oldroyd}
\frac{{\cal W}_2 }{{\cal W}_N } = \frac{\eta}{\eta_N}  \frac{1+\De_1\De_2}{1+\De_1^2},
\end{equation}
where ${\cal W}_N$ denotes the leading-order rate of working in a Newtonian fluid of viscosity $\eta_N$.

\subsection{Order $\epsilon^2$ solution and swimming velocity}\label{oldroyc_C}
We now turn to the determination of the swimming velocity of the sheet. The order $\epsilon^2$ component of  Eq.~\eqref{oldroydB} is given by 
\begin{eqnarray}\label{oldroyd_order2}
\left(1+ \De_1 \frac{\p }{\p t}\right)\btau_2 -\left(1+ \De_2 \frac{\p }{\p t}\right)\gamd_2 & = &
 \De_1(\nabla\u_1^T\cdot \btau_1+  \btau_1 \cdot \nabla\u_1 - \u_1\cdot \nabla \btau_1) \\
&& - \De_2(\nabla\u_1^T\cdot \gamd_1+  \gamd_1 \cdot \nabla\u_1 - \u_1\cdot \nabla \gamd_1).
\nonumber
\end{eqnarray}
The only relevant part of the streamfunction $\psi_2$ is its mean value, as this is the only one possibly contributing to a non-zero swimming speed in the far field. Using Fourier notations and Eq.~\eqref{stress_1}, we get the mean value of the constitutive equation, Eq.~\eqref{oldroyd_order2}, as given by 
\begin{equation}\label{oldroyd_order2_mean}
\langle \btau_2 \rangle - \langle \gamd_2 \rangle=
\Re\left\{\frac{\De_1-\De_2}{2(1+i\De_1)}
\left( \nabla\widetilde\u_1^{T*}\cdot \widetilde \gamd_1+  \widetilde\gamd_1 \cdot \nabla\widetilde\u_1^* - \widetilde\u_1^* \cdot \nabla \widetilde\gamd_1
\right)
 \right\}.
\end{equation}
The right-hand side of Eq.~\eqref{oldroyd_order2_mean} can be calculated using the leading-order kinematics, given by Eq.~\eqref{order1_sol}. We find
\begin{equation}
\widetilde\u_1=
\left[
\begin{array}{c}
 -iy     \\
 - (1+y)     
\end{array}
\right]e^{-y} e^{-ix}
,\quad
\nabla \widetilde \u_1 =
\left[
\begin{array}{cc}
-y     & i(1+y)  \\
i(y-1)  & y    
\end{array}
\right]e^{-y} e^{-ix}
,\quad 
\widetilde   \gamd_1=
\left[
\begin{array}{cc}
  -2y   &2iy  \\
2iy  &    2y 
\end{array}
\right]e^{-y} e^{-ix}
\end{equation}
leading to 
\begin{equation}\label{olroyd_average_final}
\langle \btau_2 \rangle - \langle \gamd_2 \rangle=
\left(\frac{\De_1-\De_2}{1+\De_1^2}\right)e^{-2y}
\left[
\begin{array}{cc}
  6y^2-2y-1   &\De_1(1+2y-2y^2) \\
\De_1(1+2y-2y^2)  &    2y^2+2y+1 
\end{array}
\right]\cdot
\end{equation}
We then take the divergence of Eq.~\eqref{olroyd_average_final}, and the curl to eliminate the pressure, and obtain the equation for the streamfunction
\begin{equation}\label{final_psi2}
\nabla^4 \langle \psi_2 \rangle =\frac{\De_1(\De_2-\De_1)}{1+\De_1^2}
\frac{\d ^2 }{d y^2}
\left[e^{-2y}(1+2y-2y^2)\right],
\end{equation}
with boundary conditions given by the time-average of Eq.~\eqref{bc_order2} 
\begin{subeqnarray}\label{bc_order2_average}
\frac{\p \langle \psi_2 \rangle}{\p y} ( \infty) & = & U_2,\\
\frac{\p \langle \psi_2 \rangle}{\p y} (0) & = &  \frac{1}{2}. \slabel{bc_sheet_y}
\end{subeqnarray}
The solution to Eq.~\eqref{final_psi2} is, using the boundary condition given by Eq.~\eqref{bc_sheet_y},
\begin{equation}
\frac{\d }{\d y}\langle \psi_2 \rangle (x,y) =
\frac{1}{2}\left(\frac{1+\De_1\De_2}{1+\De_1^2}\right)
+ e^{-2y}  \left(\frac{\De_1(\De_2-\De_1)}{1+\De_1^2}\right)\left(y^2-\frac{1}{2}\right),
\end{equation}
so that the swimming velocity is given by
\begin{equation}\label{velocity_oldroyd}
\frac{U_2}{U_N}=\frac{1+\De_1\De_2}{1+\De_1^2},
\end{equation}
where $U_N=1/2$ is the leading-order  swimming velocity obtained in a Newtonian fluid.
The velocity of the free sheet (or, alternatively, the flow speed induced in the far-field for a fixed sheet) is therefore always smaller than the Newtonian value ($U \leq U_N$).

\section{High Deborah numbers: FENE-P}
\label{FENE}

As was discussed above, the Oldroyd-B constitutive equation is a very attractive model for polymeric fluids, but it is not valid for large extension rates. In fact, in a purely extensional flow, its extensional viscosity blows up at a finite value of the extension rate. For the setup considered in this paper, the flow near the sheet is a combination of shear and extensional flow, and we are therefore not allowed to use the results obtained in the previous section, Eqs.~\eqref{work_oldroyd} and \eqref{velocity_oldroyd}, in the biologically-relevant limit of large Deborah numbers. In order to remedy this issue, we consider in this section a more advanced constitutive relationship, the FENE-P model \cite{bird80,birdvol1,birdvol2,larson88,tanner88,bird95,larson99}. The physical picture of the FENE-P model is very similar to Oldroyd-B, with the difference that the entropic springs modeling polymer molecules have finite extension and can therefore sustain arbitrarily large strain rates \cite{bird80,birdvol1,birdvol2,larson88,tanner88,bird95,larson99}. The model can be exactly derived for a solution of finite-extension dumbbells  expect for one empirical averaging step (the Peterlin approximation  \cite{herrchen97,keunings97}, at the origin of the ``P'' in FENE-P).

\subsection{Model}
In the FENE-P model, the polymeric stress is given by 
\begin{equation}\label{FENE_a}
\btau^p =G\left[f(\bka) \bka- {\bf 1}\right],
\end{equation}
where $G$ has units of stress, ${\bf 1}$ is the identity tensor, and $\bka$ is the dimensionless dumbbell  covariance tensor, which satisfies the evolution equation
\begin{equation}\label{FENE_b}
f(\bka) \bka + \lambda  {\stackrel{\triangledown}{\bka}} = \I,
\end{equation}
where  $\lambda$ is the (single) polymer relaxation time. The function $f$ is  defined as
\begin{equation}\label{FENE_c}
f(\bka)=\frac{1}{1-\frac{\tr (\bka)}{L^2}},
\end{equation}
where $L$ is a (large) dimensionless number measuring the maximum relative extension of the polymer molecules. In the limit where $L\to\infty$, this model reduces to the  Oldroyd-B  model, Eq.~\eqref{UCM}, with $\eta_p = G \lambda$.

As in the previous section, shear rates and stresses are nondimensionalized by $\omega$ and $\eta \omega$ respectively, and the value of the total viscosity, $\eta$, will be defined below. All symbols will now refer to dimensionless variables.

\subsection{Expansion}
At equilibrium, we obtain the order $\epsilon^0$  solution, with no flow ($\u_0=0$) and 
\begin{equation}\label{FENE_eq}
\bka_0 = \frac{1}{1+\frac{3}{L^2}}\I,\quad f(\bka_0) = f_0 = 1+\frac{3}{L^2},\quad 
\btau_0^p = 0.
\end{equation}
We then look for a regular perturbation expansion in the form
\begin{equation}\label{exp_kappa}
\bka - \bka_0 = \epsilon \bka_1 + \epsilon^2 \bka_2 + ...,
\end{equation}
and therefore
\begin{equation}
f(\bka) - f_0 = \epsilon f_1 + \epsilon^2 f_2.
\end{equation}
Using Eq.~\eqref{exp_kappa}, we see that
\begin{equation}
f_1= \frac{(3+L^2)^2}{L^6} \tr (\bka_1)
,\quad
f_2 =\frac{(3+L^2)^2}{L^6} \tr (\bka_2) + \frac{(3+L^2)^3}{L^{10}} \tr (\bka_1).
\end{equation}

\subsection{Order $\epsilon$ solution and rate of work}
At order $\epsilon$,  we have
\begin{equation}\label{FENE_order1} 
f_1 \bka_0 + f_0 \bka_1 + \lambda \omega \frac{\p \bka_1}{\p t} = \lambda \omega
[\nabla \u_1^T\cdot \bka_0 + \bka_0 \cdot \nabla \u_1- \u_1 \cdot \nabla \bka_0],
\end{equation}
and the last term on the right-hand side of Eq.~\eqref{FENE_order1} is zero as $\bka_0$ is constant. 
Using the fact that $\bka_0=\I/f_0$, 
Eq.~\eqref{FENE_order1} becomes
\begin{equation}\label{FENE_order1_new}
f_1 \bka_0 + f_0 \bka_1 + \lambda \omega \frac{\p \bka_1}{\p t} = \frac{\lambda\omega }{f_0}
\gamd_1,
\end{equation}
and by taking the trace of Eq.~\eqref{FENE_order1_new} we find $\tr \bka_1=f_1=0$. It is then natural to  define the first Deborah number as
\begin{equation}\label{newDe}
\De_1=\frac{\lambda \omega}{f_0},
\end{equation}
and since $(f_0 - 1) \ll 1$, this definition is essentially equivalent to the definition 
$\De_1= \lambda \omega$ for  the Oldroyd-B model.
We then  obtain, using Fourier space notations, 
\begin{equation}\label{kba_1_fourier}
\widetilde \bka_1 = \frac{1}{f_0}\frac{\De_1}{1+ i \De_1}\widetilde\gamd_1\cdot
\end{equation}
The equation for the dimensionless polymeric stress is
$\eta \omega \btau^p_1 = G  f_0 \bka_1 $,
so that the total stress,  $\btau_1 = \btau^s_1 + \btau^p_1$, is given in Fourier space by
\begin{equation}
 \eta \omega\widetilde \btau_1 = \eta_s  \omega \widetilde\gamd_1 + 
G \frac{\De_1}{1+ i \De_1}\widetilde\gamd_1\cdot
\end{equation}
In the limit of small Deborah number $\De_1$, the expression above allows us to define without ambiguity the polymer viscosity, and therefore the total viscosity and second Deborah number, which are given by 
\begin{equation}\label{newetap}
\eta_p = \frac{G\lambda}{f_0}
,\quad 
\eta = \eta_s + \eta_p
,\quad
\De_2  = \De_1 \frac{\eta_s}{\eta}\cdot
\end{equation}
The first-order non-dimensional stress is finally by
\begin{equation}
\widetilde \btau_1 = \frac{1+i\De_2}{1+ i \De_1}\widetilde\gamd_1.
\end{equation}
This equation is identical to the one obtained for the Oldroyd-B model, Eq.~\eqref{stress_1}, and therefore we find that, provided that the Deborah number and polymer viscosity are  defined using Eqs.~\eqref{newDe} and \eqref{newetap}, the leading-order flow kinematics as well as the rate of work of the sheet are the same as for the Newtonian case, and given by Eqs.~\eqref{order1_sol} and \eqref{work_oldroyd}.

\subsection{Order $\epsilon^2$ solution and swimming velocity}
At next order, the evolution equation for $\bka$ becomes
\begin{equation}\label{FENE2_a}
f_2 \bka_0 + f_0 \bka_2 + f_0 \De_1 \frac{\p \bka_2}{\p t} =f_0 \De_1
[\nabla \u_2^T\cdot \bka_0 + \bka_0 \cdot \nabla \u_2 
+
\nabla \u_1^T\cdot \bka_1 + \bka_1 \cdot \nabla \u_1 - u_1\cdot \nabla \bka_1],
\end{equation}
and the polymer stress is given by
\begin{equation}\label{FENE2_b}
\btau^p_2 = \left(\frac{\De_1-\De_2}{\De_1^2}\right) \left[ f_2 \bka_0 + f_0 \bka_2 \right],
\end{equation}
where have used that $G/\eta \omega = (\De_1-\De_2)/\De_1^2$. Combining Eqs.~\eqref{FENE2_a} and \eqref{FENE2_b}, and exploiting the fact that $\bka_0=\I/f_0$, we obtain the expression for the average of the second-order stress as
\begin{equation}
\langle \btau^p_2 \rangle = \left( 1- \frac{\De_2}{\De_1}\right) \langle \gamd_2 \rangle 
+ f_0 \left( 1- \frac{\De_2}{\De_1}\right)  \langle  \nabla \u_1^T\cdot \bka_1 + \bka_1 \cdot \nabla \u_1 - u_1\cdot \nabla \bka_1\rangle,
\end{equation}
and therefore, with $\btau_2 = \btau^s_2 + \btau^p_2$, we obtain
\begin{equation}
\langle \btau_2 \rangle -   \langle \gamd_2 \rangle 
= f_0 \left( 1- \frac{\De_2}{\De_1}\right)  \langle  \nabla \u_1^T\cdot \bka_1 + \bka_1 \cdot \nabla \u_1 - u_1\cdot \nabla \bka_1\rangle.
\end{equation}
Using Fourier notations we find 
\begin{equation}
\langle \btau_2 \rangle -   \langle \gamd_2 \rangle 
=\Re\left\{ 
  \frac{f_0}{2} \left( 1- \frac{\De_2}{\De_1}\right) 
  \left( \nabla\widetilde\u_1^{T*}\cdot \widetilde \bka_1+  \widetilde\bka_1 \cdot \nabla\widetilde\u_1^* - \widetilde\u_1^* \cdot \nabla \widetilde\bka_1
\right)
\right\},
\end{equation}
so that, exploiting the result of Eq.~\eqref{kba_1_fourier}, we obtain
\begin{equation}
\langle \btau_2 \rangle -   \langle \gamd_2 \rangle 
= \Re\left\{ \frac{\De_1-\De_2}{2(1+i\De)}
 \left( \nabla\widetilde\u_1^{T*}\cdot \widetilde \gamd_1+  \widetilde\gamd_1 \cdot \nabla\widetilde\u_1^* - \widetilde\u_1^* \cdot \nabla \widetilde\gamd_1
\right)
\right\}.
\end{equation}
This expression is identical to the one obtained  for the Oldroyd-B model, Eq.~\eqref{oldroyd_order2_mean}, and therefore the swimming speed of the sheet is the same as well, given by Eq.~\eqref{velocity_oldroyd}.

\section{Discussion and implications for biological transport}
\label{discussion}

In this paper, we have shown that the presence of a non-Newtonian fluid modifies significantly the kinematics and energetics of transport and locomotion by a waving sheet as compared to that in a Newtonian fluid. Specifically, we have demonstrated that the leading-order swimming velocity of a free sheet (or, alternatively, the induced far-field velocity field when the sheet is fixed) is given by 
\begin{equation}\label{finalU}
\frac{U}{U_N} = \frac{1+\De^2{\eta_s}/{\eta}}{1+\De^2},
\end{equation}
where 
\begin{equation}\label{UN}
U_N = \frac{1}{2}\omega a^2 k,
\end{equation}
is the leading-order (dimensional) velocity obtained in a Newtonian fluid;  next order terms in Eq.~\eqref{UN} are  $O(\omega a^4k^3)$. 
With the appropriate definition of the Deborah number and the polymer viscosity, this result was shown to be valid for two constitutive relationships,  Oldroyd-B (\S\ref{oldroyd}) and FENE-P (\S\ref{FENE}), but other models such as Johnson-Segalman-Oldroyd (see Appendix \ref{JSO}) and Giesekus (see Appendix \ref{G}) lead the same result for the swimming speed. Furthermore, as shown in Appendix \ref{tangential}, the result is also unchanged if the wave of displacement propagating along the sheet  includes not only normal but also tangential motion. The generalization of Eq.~\eqref{finalU} to the case of many relaxation times is presented in Appendix \ref{manymodes}.

Since the zero-shear-rate viscosity of the polymeric fluid, $\eta = \eta_s + \eta_p $, is larger than the viscosity of the Newtonian solvent, $\eta_s$ (and in most cases, much larger, see \S\ref{intro}), we always have $U \leq U_N$. Locomotion with a given gait is therefore always slower in a non-Newtonian fluid. In the biologically-relevant limit of large Deborah numbers, the velocity  asymptotes to a constant value, equal to $U_N \eta_s/\eta$, and therefore the swimming speed is significantly reduced from the Newtonian value.  The dimensional dependance on the parameters of the swimming gait,  Eq.~\eqref{UN}, suggests that such reduction in velocity could be compensated by  an increase in the beat frequency (still consistent with the $\De\gg1 $ limit) and a decrease in the wavelength, both of which are observed for spermatozoa motion in cervical mucus \cite{katz78}. An increase in the beat amplitude would also increase the swimming velocity, however  sperm cells are observed to swim in cervical mucus with decreased amplitude \cite{katz78}.

The leading-order rate of work done by the waving sheet, as compared to the work ${\cal W}_N$ in a Newtonian fluid of viscosity $\eta_N $, is given by 
\begin{equation}\label{finalW}
\frac{{\cal W}}{{\cal W}_N} = \frac{\eta}{\eta_N}  \frac{1+\De^2{\eta_s}/{\eta}}{1+\De^2} \cdot
\end{equation}
Notably, this result is identical to that given by forced oscillations of the  Jeffreys spring-dashpot model: a dashpot of viscosity $\eta_s$ in a parallel with a Maxwell model, {\it i.e.} of a dashpot of viscosity $\eta_p$ in series with a spring of constant $k_p = \eta_p/\lambda$.

It is reasonable to consider two different cases when discussing Eq.~\eqref{finalW}. Firstly, we can compare the motion in the polymeric fluid with the motion in the solvent alone, which requires  a rate of  work ${\cal W}_s$. In that case, we need to pick $\eta_N = \eta_s$, and Eq.~\eqref{finalW} becomes 
\begin{equation}
\frac{{\cal W} }{{\cal W}_s } =\frac{ {\eta}/{\eta_s}+\De^2}{1+\De^2}\cdot
\end{equation}
Swimming in a polymeric fluid requires therefore always more work than swimming in the solvent alone (${\cal W} \geq  {\cal W}_s$); note however that in the limit of large Deborah numbers, we get ${\cal W} \approx {\cal W}_s$. We can also compare the rate of viscous dissipation with that occurring in the Newtonian fluid which has the same zero-shear-rate viscosity as the polymeric fluid, with rate of work ${\cal W}_{\eta}$. In that case, we choose $\eta_N = \eta$ and obtain
\begin{equation}
\frac{{\cal W} }{{\cal W}_\eta } =\frac{1+\De^2{\eta_s}/{\eta}}{1+\De^2}\cdot
\end{equation}
As a difference, we see here that compared to a Newtonian fluid with the same viscosity, it is energetically advantageous to swim in a fluid that has some elasticity (${\cal W} \leq  {\cal W}_\eta$).

The mechanical efficiency of the swimming motion, defined as the ratio of useful work to total work, is given by
\begin{equation}\label{eff}
{\cal E}\sim \frac{\eta U^2 }{{\cal W}},
\end{equation}
where the units in Eq.~\eqref{eff} are correct as ${\cal W}$ is a rate of work per unit length in the spanwise   direction ($z$ direction, see Fig.~\ref{mainfig}). Given Eqs.~\eqref{finalU} and \eqref{finalW}, we obtain
\begin{equation}\label{finalE}
\frac{{\cal E}}{{\cal E}_N} =  \frac{1+\De^2{\eta_s}/{\eta}}{1+\De^2},
\end{equation}
when ${\cal E}_N$ is the efficiency for the swimming motion in the Newtonian fluid. Again, we obtain ${\cal E} \leq {\cal E}_N$ and swimming in a non-Newtonian fluid deteriorates the mechanical efficiency.

At first, these observations seem to suggest that there are only disadvantages to swimming in a complex fluid, as compared to a Newtonian fluid. However, we would like to argue here that non-Newtonian fluids could be exploited by biological systems to their advantage because viscoelastic stresses allow a tuning of the kinematics of transport and locomotion in a manner which is not possible with a Newtonian fluid. Indeed, in the case of a Newtonian fluid, the  swimming velocity of the sheet is given by Eq.~\eqref{UN}, and it is only a function of the swimming gait parameters. The only way to modify, for example, the speed at which cilia transport a Newtonian mucus would be to change the amplitude, frequency or wavelength of the ciliary waves, and a viscosity change would only impact  the necessary rate of work. However, as Eq.~\eqref{finalU} shows, the transport kinematics in a viscoelastic fluid are also controlled by the mechanical properties of the fluid, namely its viscosity and relaxation time. This new feature allows the transport to be tuned without having to modify the parameters of the swimming gait, a feature which is notably not present when the fluid is Newtonian and could potentially be important. 

This observation suggests the possibility, for a biological system, to modulate transport and motility passively. For example, tuning the rheology of respiratory mucus would influence the rate at which it is transported and evacuated without having to control the cilia themselves. Similarly, varying the rheological properties of the cervical mucus would allow the female reproductive tract to select appropriately motile spermatozoa without requiring a modification of the  waveforms and beat frequency of their flagella. A number of experimental observations support this idea. The rheological properties of cervical mucus vary during the menstrual cycle, and measured  ``sperm penetrability'' peaks during the ovulatory phase when the viscosity of the mucus is at a minimum  \cite{wolf77_3,wolf78,james83}. This is consistent with our result ${\p U}/{\p \eta_p} < 0$, easily inferred from Eq.~\eqref{finalU}. The elasticity of the fluid has also been reported to be an important factor \cite{king74,gelman79,litt81} and transport usually decreases with an increase in the elasticity \cite{meyer76,giordano78}, an observation which  is consistent with our result ${\p U}/{\p \De} < 0$ (see Eq.~\ref{finalU}) \footnote{Note that, from Eqs.~\eqref{finalW} and \eqref{finalE}, we also see that ${\p {\cal W}}/{\p \De} < 0$ and $ {\p {\cal E}}/{\p \De} < 0 $.}. Similar results were obtained 
in numerical simulation of mucociliary transport \cite{ross74} and spermatozoa motion in deformable domains \cite{fauci95}.

In conclusion, we have revisited in this paper Taylor's swimming sheet calculation in the case where the fluid is non-Newtonian. Although it is clearly a very simplified model, the results obtained in this paper could be extended to more complicated geometries and wave profiles, and could therefore provide a first step towards a rigorous quantitative approach to the energetics and kinematics of locomotion and transport in complex  biological fluids.


\appendix
\section{Retarded-motion expansion}
\label{retarded}

In this section, we re-derive the result first obtained in Ref.~\cite{chaudhury79} and show that a waving sheet in a second-order fluid swims at the same speed as in a Newtonian fluid. We also show that the rate of working of the sheet is unchanged.

A second-order fluid is the first term in a systematic asymptotic expansion of the relationship 
between the stress and the rate-of-strain tensor for small  Deborah numbers and is valid in the limit of slow and slowly varying flows  \cite{bird76,birdvol1,birdvol2,larson88,tanner88,bird95}. The instantaneous relationship is given by
\begin{equation}
\btau = \eta_0 \gamd - \frac{1}{2}\Psi_1 \dgamd + \Psi_2 (\gamd\cdot\gamd) ,
\end{equation}
where $\Psi_1$ and $\Psi_2$ are the first and second normal stress coefficients. We nondimensionalize stresses by $\eta_0 \omega$, shear rates by $\omega$, and the relationship becomes
\begin{equation}\label{rme}
\btau =  \gamd - \De_1 \dgamd - \De_2 (\gamd\cdot\gamd),
\end{equation}
where $\De_1 = \Psi_1 \omega / 2 \eta_0$ and $\De_2 =- \Psi_2 \omega / \eta_0$ are the two characteristic Deborah numbers (both are expected to be positive).

At first order, Eq.~\eqref{rme} becomes
\begin{equation}
\btau_1 = \left(1 - \De_1 \frac{\p }{\p t}\right)\gamd_1,
\end{equation}
and therefore the streamfunction satisfies
\begin{equation}
\left(1 - \De_1 \frac{\p }{\p t}\right)\nabla^4 \psi_1 = 0.
\end{equation}
Similarly to the result of \S\ref{oldroyd_order1}, we see that the solution is the same as that obtained by Taylor for motion in a Newtonian fluid as given by Eq.~\eqref{taylor_1}.
The first-order component of the stress tensor is given by
\begin{equation}
\widetilde \btau_1 = (1-i\De_1) \widetilde \gamd_1 ,
\end{equation}
and therefore the rate of work is equal to the volume integral of 
\begin{equation}
\langle \btau_1:\gamd_1 \rangle = \frac{1}{2}\Re\{ \widetilde\btau_1: \widetilde\gamd_1^*\}= \widetilde\gamd_1: \widetilde\gamd_1^*,
\end{equation}
and is the same as for the case of a Newtonian fluid (with equal viscosity $\eta_0$).

At second order,  Eq.~\eqref{rme} is given by
\begin{equation}\label{rme_2}
\btau_2 -  \left(1 - \De_1 \frac{\p }{\p t}\right) \gamd_2 = \De_1 (\nabla\u_1^T\cdot \gamd_1+  \gamd_1 \cdot \nabla\u_1 - \u_1\cdot \nabla \gamd_1) - \De_2 (\gamd_1\cdot\gamd_1),
\end{equation}
and its mean value is given by, using Fourier notations,
\begin{equation}\label{rme_2_average}
\langle \btau_2 \rangle  -  \langle  \gamd_2 \rangle  = 
\frac{1}{2}\Re\left\{\De_1
\left( \nabla\widetilde\u_1^{T*}\cdot \widetilde \gamd_1+  \widetilde\gamd_1 \cdot \nabla\widetilde\u_1^* - \widetilde\u_1^* \cdot \nabla \widetilde\gamd_1
\right) -\De_2(\widetilde \gamd_1\cdot \widetilde\gamd_1^*)
 \right\}.
\end{equation}
Using Eq.~\eqref{order1_sol},  we get
\begin{equation}
\langle \btau_2 \rangle  -  \langle  \gamd_2 \rangle  = 
e^{-2y}\left[
\begin{array}{cc}
 (6y^2-2y-1)\De_1-4y^2\De_2   &0\\
0  &    (2y^2+2y+1)\De_1 - 4y^2\De_2
\end{array}
\right], 
\end{equation}
for which we can take the divergence and then the curl to obtain
\begin{equation}
\nabla^4 \langle \psi_2 \rangle =0.
\end{equation}
The equation and the boundary conditions are the same as for the Newtonian problem, and therefore the swimming velocity is unchanged.

\section{Johnson-Segalman-Oldroyd}
\label{JSO}

The Johnson-Segalman-Oldroyd constitutive relationship is an empirical model capturing non-affine motions in the polymeric fluid \cite{johnson77}. It is formally similar to the Oldroyd-B constitutive relationship but uses the (more general) Gordon-Schwalter convected derivative. It includes other simple models such as Oldroyd-B, Oldroyd-A, and corotational Maxwell as special cases \cite{birdvol1,birdvol2,larson88}.

The constitutive relationship is given by 
\begin{equation} \label{JSO_eq}
\btau +\lambda_1 \dsbtau= \eta [\gamd + \lambda_2 \dsgamd ],
\end{equation}
where the Gordon-Schwalter  convected derivative is 
\begin{equation}\label{JSO_newderiv}
\dsbtau = \dbtau + \frac{a}{2} (\gamd \cdot \btau + \btau\cdot \gamd ),
\end{equation}
and where $a$ is a dimensionless parameter related to second normal stress differences and leading to shear-thinning behavior in steady shear.
Using the same non-dimensionalization as for Oldroyd-B, we obtain the dimensionless constitutive relationship
\begin{equation}\label{JSO_newderiv_NODIM}
\btau +\De_1 \dsbtau = \gamd + \De_2 \dsgamd.
\end{equation}
Since the linear part of Eq.~\eqref{JSO_newderiv_NODIM} is the same as for Oldroyd-B, it is straightforward to see that the problem at order $\epsilon$ is also the same as for Oldroyd-B, so the rate of working of the sheet is given by Eq.~\eqref{finalW}.  The change in the convected derivative will affect at most the second-order solution. The constitutive equation at second order has two terms added as a result of the change in the convected derivatives 
\begin{eqnarray}\label{JS_order2}
\left(1+ \De_1 \frac{\p }{\p t}\right)\btau_2 -\left(1+ \De_2 \frac{\p }{\p t}\right)\gamd_2 & = & 
 \De_1(\nabla\u_1^T\cdot \btau_1+  \btau_1 \cdot \nabla\u_1 - \u_1\cdot \nabla \btau_1)\\ 
 &&-\frac{a}{2}\De_1(\gamd_1\cdot \btau_1 + \btau_1 \cdot \gamd_1 ) \nonumber\\
 && 
- \De_2(\nabla\u_1^T\cdot \gamd_1+  \gamd_1 \cdot \nabla\u_1 - \u_1\cdot \nabla \gamd_1) \nonumber\\
&& + a\De_2(\gamd_1\cdot \gamd_1),
\nonumber
\end{eqnarray}
which leads to the modified equation for the mean stress
\begin{eqnarray}\label{JS_order2_mean}
\langle \btau_2 \rangle - \langle \gamd_2 \rangle&=&
\Re\left\{\frac{\De_1-\De_2}{2(1+i\De_1)}
\left( \nabla\widetilde\u_1^{T*}\cdot \widetilde \gamd_1+  \widetilde\gamd_1 \cdot \nabla\widetilde\u_1^* - \widetilde\u_1^* \cdot \nabla \widetilde\gamd_1 
\right)
 \right\}\\
&& + \frac{a}{4}\left(\frac{\De_2-\De_1}{1+\De_1^2}\right)(\widetilde \gamd_1 \cdot \widetilde \gamd_1^* + \widetilde \gamd_1^* \cdot \widetilde \gamd_1)\nonumber.
\end{eqnarray}
The new term on the right-hand side of Eq.~\eqref{JS_order2_mean} is given by
\begin{equation}\label{RHS}
\widetilde \gamd_1 \cdot \widetilde \gamd_1^* + \widetilde \gamd_1^* \cdot \widetilde \gamd_1 = 16 y^2 e^{-2y}
{\bf 1}.
\end{equation}
The tensor given in Eq.~\eqref{RHS} is proportional to the identity tensor, ${\bf 1}$, and can therefore be absorbed in the definition of the fluid pressure. Consequently, this new term does not modify the equation for the streamfunction as given by Eq.~\eqref{final_psi2} and the swimming velocity is the same as for Oldroyd-B (Eq.~\ref{finalU}). All Oldroyd-like models lead therefore to the same results.

\section{Giesekus}
\label{G}

As our last model, we consider the Giesekus constitutive relationship \cite{giesekus82}, which, although empirical, was derived using molecular ideas for polymer networks  \cite{bird76,birdvol1,birdvol2,larson88,tanner88,bird95,larson99} and is therefore directly relevant to complex biological fluids. In this model, the polymeric contribution to the stress satisfies the evolution equation
\begin{equation}
\btau^p + \lambda  {\stackrel{\triangledown}{\btau^p}} + \alpha \frac{\lambda}{\eta_p} (\btau^p \cdot \btau^p) = \eta_p \gamd,
\end{equation}
where $\alpha$ is a small dimensionless positive number related to the second normal stress coefficient in steady shear.
The equation for the total stress $\btau$ is then found to be
\begin{equation}\label{Giesekus_final}
\btau + \lambda_1\left[  {\stackrel{\triangledown}{\btau}} 
-\alpha\frac{\eta_s}{\eta_p}  (\gamd \cdot \btau + \btau\cdot \gamd ) 
\right]
+
\alpha \frac{\lambda_1}{\eta_p} (\btau \cdot \btau)
=\eta \left[
 \gamd + \lambda_2 \left[  {\stackrel{\triangledown}{\gamd}} -\alpha \frac{\eta_s}{\eta_p}(\gamd \cdot \gamd)\right]
\right], 
\end{equation}
where, again, $\eta= \eta_s+\eta_p$ and $\lambda_1=\lambda$ and $\lambda_2= \lambda \eta_s / \eta$. Compared to the Oldroyd-B constitutive equation, three new quadratic terms appear. The new terms of the from  $(\gamd\cdot \gamd)$ and $ (\gamd \cdot \btau)$ are formally similar to the extra terms appearing in the Johnson-Segalman-Oldroyd constitutive relationship, Eqs.~\eqref{JSO_eq} and \eqref{JSO_newderiv}, and do not modify the swimming velocity (see Appendix \ref{JSO}). We therefore only need to quantify the influence of the term quadratic in the stress, last term on the left-hand side of Eq.~\eqref{Giesekus_final}. In non dimensional and complex notations, we still have at first order
\begin{equation}\widetilde \btau_1 = \frac{1+i\De_2}{1+i\De_1} \widetilde \gamd_1 ,
\end{equation}
so that the expression for the rate of work is unchanged, and the steady contribution to the second-order streamfunction arises from the new term
\begin{equation}
\langle \btau_1\cdot \btau_1 \rangle = 
\frac{1}{4}(\widetilde \btau_1 \cdot \widetilde \btau_1 ^* + \widetilde \btau_1 ^* \cdot \widetilde \btau_1 
)
=\frac{1}{4}\left(\frac{1+\De_2^2}{1+\De_1^2}\right)(\widetilde \gamd_1 \cdot \widetilde \gamd_1^* + \widetilde \gamd_1^* \cdot \widetilde \gamd_1).
\end{equation}
As for the Johnson-Segalman-Oldroyd model, this term does not modify the equation for the mean value of the order-two streamfunction, and therefore does not impact the swimming kinematics. As a consequence, our final results, Eqs.~\eqref{finalU} and \eqref{finalW}, are also valid for the Giesekus model.

\section{Allowing tangential motion of the sheet}
\label{tangential}

Here, we show that the main results of this paper, Eqs.~\eqref{finalU} and \eqref{finalW}, are unchanged is we allow the wave of displacement along the sheet to include not only normal but also tangential motion.  In the reference frame moving with the sheet, we write the dimensionless position of material points, $(x_m,y_m)$ on the sheet as
\begin{equation}
x_m =  x  +\epsilon  a \cos(x-t - \phi)
,\quad
y_m=\epsilon b \sin (x - t) ,
\end{equation}
where  $a$ and $b$ are dimensionless and $\phi$ is the phase difference between the tangential and normal motion. The boundary condition in the far-field is unchanged,
\begin{equation}
\nabla \psi \big\vert_{(x_m,\infty)}  = U\ey,
\end{equation}
whereas the boundary conditions on the sheet are modified and given by
\begin{equation}
\nabla \psi\big\vert_{(x_m,y_m)}   = \epsilon (b\cos (x-t) \ex + a \sin (x-t - \phi) \ey).
\end{equation}
At order $\epsilon$, they become
\begin{subeqnarray}
\nabla \psi_1 \big\vert_{(x,\infty)}  & = & U_1 \ey,  \\
\nabla \psi_1 \big\vert_{(x,0)} & = &  b\cos (x-t) \ex + a \sin (x-t - \phi) \ey,
\end{subeqnarray}
and at order $\epsilon^2$ we obtain
\begin{subeqnarray}
\nabla \psi_2 \big\vert_{(x,\infty)}  & = &  U_2\ey,\\
\nabla \psi_2 \big\vert_{(x,0)} & = & - a \cos (x-t - \phi) \nabla \left(\frac{\p \psi_1}{\p x}\right) \bigg\vert_{(x,0)}
-b \sin(x-t) \nabla \left(\frac{\p \psi_1}{\p y}\right) \bigg\vert_{(x,0)}.
\end{subeqnarray}
For the Oldroyd-B model, the problem at order $\epsilon$ is straightforward to solve and we obtain 
\begin{equation}
\tp_1=i[b+(b+ae^{i\phi})y]e^{-y}e^{-ix} =
ib[1+(1+c)y]e^{-y}e^{-ix} ,
\end{equation}
where we have defined, $c=ae^{i\phi}/b$.  Since the equation quantifying the mean value of the rate of work,  Eq.~\eqref{mean_work}, is not modified, the final formula for the rate of work of the sheet is unchanged, and given by Eq.~\eqref{finalW}.
We can then compute
\begin{equation}
\widetilde\u_1=
b\left[
\begin{array}{c}
 i c - i(1 + c)y  \\
 - 1-(1+c)y     
\end{array}
\right]e^{-y} e^{-ix},
\end{equation}
\begin{equation}
\nabla \widetilde \u_1 =
b\left[
\begin{array}{cc}
c - (1+c)y     & i   + i (1+c)y  \\
-i(1+2c) + i(1+c)y  & -c + (1+c)y    
\end{array}
\right]e^{-y} e^{-ix}
,
\end{equation}
\begin{equation}
\nabla \widetilde \u_1^* =
b\left[
\begin{array}{cc}
c^* - (1+c^*)y     & -i   - i (1+c^*)y  \\
i(1+2c^*) - i(1+c^*)y  & -c^* + (1+c^*)y    
\end{array}
\right]e^{-y} e^{ix}
,
\end{equation}
\begin{equation}
\nabla \widetilde \u_1^{T*} =
b \left[
\begin{array}{cc}
c^* - (1+c^*)y     & i(1+2c^*) - i(1+c^*)y\\
 -i   - i (1+c^*)y    & -c^* + (1+c^*)y    
\end{array}
\right]e^{-y} e^{ix}
,
\end{equation}
and
\begin{equation}
\widetilde   \gamd_1=
b\left[
\begin{array}{cc}
  2c - 2(1+c)y    &-2ic + 2i(1+c)y  \\
-2ic + 2i(1+c)y  &    -2c + 2(1+c)y
\end{array}
\right]e^{-y} e^{-ix}.
\end{equation}
Using the previously-derived relationship
\begin{equation}
\langle \btau_2 \rangle - \langle \gamd_2 \rangle=
\Re\left\{\frac{\De_1-\De_2}{2(1+i\De_1)}
\left( \nabla\widetilde\u_1^{T*}\cdot \widetilde \gamd_1+  \widetilde\gamd_1 \cdot \nabla\widetilde\u_1^* - \widetilde\u_1^* \cdot \nabla \widetilde\gamd_1
\right)
 \right\},
\end{equation}
we obtain, after some algebra, 
\begin{equation}\label{neworder4}
\nabla^4 \langle \psi_2 \rangle = \frac{\De_1(\De_2-\De_1)}{1+\De_1^2}b^2
\frac{\d ^2 }{d y^2}
\left[e^{-2y}(1-3cc^*+2y(1+2c+2c^*+3cc^*)-2y^2(1+c)(1+c^*))\right],
\end{equation}
associated with the boundary conditions
\begin{subeqnarray}
\frac{\p \langle \psi_2 \rangle}{\p y} ( \infty) & = & U_2,\\
\frac{\p \langle \psi_2 \rangle}{\p y} (0) & = &  \frac{1}{2}(b^2 + 2ab \cos \phi -a^2).
\end{subeqnarray}
Eq.~\eqref{neworder4} is easily integrated three times and we obtain
\begin{equation}
U_2= \frac{1}{2}(b^2+2ab\cos\phi -a^2) + \frac{1}{2}\frac{\De_1(\De_2-\De_1)}{1+\De_1^2}b^2(1+c+c^*-cc^*),
\end{equation}
and since $b^2(1+c+c^*-cc^*) = (b^2 + 2ab \cos\phi - a^2)$, we get
\begin{equation}
\frac{U_2}{U_N}= \frac{1+\De_1\De_2}{1+\De_1^2},
\end{equation}
where $U_N$ is the swimming velocity in a Newtonian fluid,
\begin{equation}
U_N =  \frac{1}{2}(b^2+2ab\cos\phi -a^2).
\end{equation}
Allowing tangential motion does therefore reduce the swimming speed by the same amount, and the final results of the papers, Eq.~\eqref{finalU} and \eqref{finalW}, remain also valid in this case.

\section{Multimode Oldroyd B model}
\label{manymodes}
In this section, we generalize the results of the paper to the case where the constitutive relationship for the viscoelastic fluid possesses more than one relaxation times \cite{birdvol1,birdvol2,larson88,tanner88,larson99}.
In that case, the polymeric contribution to the stress tensor is given by
\begin{equation}
\btau^p = \sum_{j=1}^N \btau^p_j,
\end{equation}
where each mode satisfies an upper-convected Maxwell equation
\begin{equation}\label{eachorder}
\btau^p_j + \lambda_j  {\stackrel{\triangledown}{\btau^p_j}} = \eta_j \gamd,
\end{equation}
characterizes by a relaxation time $\lambda_j$ and viscosity $\eta_j$. At low shear rates, the fluid has a steady-shear viscosity given by 
\begin{equation}
\eta= \eta_s + \sum_{j=1}^N \eta_j.
\end{equation}
We now nondimensionalize stresses by $\eta \omega$, and shear rates by $\omega$, and solve for the flow order by order. Using Fourier  transforms, we obtain, in dimensionless variables, 
\begin{equation}\label{eachorder_fourier}
\widetilde \btau^p_{j,1} = \frac{\eta_j}{\eta}\frac{1}{1+i\De_j} \widetilde \gamd_1,
\end{equation}
where $\De_j = \lambda_j \omega$. The total stress becomes therefore
\begin{equation}
\widetilde \btau _1 = \widetilde \btau^s_1 + \sum_{j=1}^N \widetilde \btau^p_{j,1} = \left( \frac{\eta_s}{\eta } + \sum_{j=1}^N \frac{\eta_j}{\eta}\frac{1}{1+i\De_j}\right) \widetilde \gamd_1,
\end{equation}
and the leading-order kinematics are unchanged from that obtained in the case of a Newtonian fluid, as given by Eq.~\eqref{order1_sol}. The rate of work done by the sheet is equal to the volume integral of 
\begin{equation}
\langle \btau_1:\gamd_1 \rangle = \frac{1}{2}\Re\{ \widetilde\btau_1: \widetilde\gamd_1^*\}=
\frac{1}{2}\Re \left\{  \frac{\eta_s}{\eta } + \sum_{j=1}^N \frac{\eta_j}{\eta}\frac{1}{1+i\De_j} \right\}
[ \widetilde\gamd_1: \widetilde\gamd_1^*],
\end{equation}
and therefore we obtain
\begin{equation}\label{finalW_multimode}
\frac{\W}{\W_N} = \frac{1}{\eta_N}  \left( {\eta_s}+ \sum_{j=1}^N {\eta_j} \frac{1}{1+\De_j^2}\right) 
= \frac{\eta}{\eta_N}  \left( 1-  \sum_{j=1}^N \frac{\eta_j}{\eta} \frac{\De_j^2}{1+\De_j^2}\right) ,
\end{equation}
where, as above, $\W_N$ is the rate of work in a Newtonian fluid of viscosity $\eta_N$. In the case of a single relaxation time scale, Eq.~\eqref{finalW_multimode} is the same as  Eq.~\eqref{finalW} with $\eta_1 = \eta_p  = \eta-\eta_s$ and $\De_1=\De$.

At next order in $\epsilon$, the mean value of the dimensionless form of  Eq~\eqref{eachorder} leads to 
\begin{equation}
\langle \btau^p_{j,2} \rangle  - \frac{\eta_j}{\eta} \langle \gamd_2 \rangle  =
 \De_j\langle \nabla\u_1^T\cdot \btau^p_{j,1}+  \btau^p_{j,1} \cdot \nabla\u_1 - \u_1\cdot \nabla \btau^p_{j,1}\rangle ,
\end{equation}
which becomes, in Fourier space, and using Eq.~\eqref{eachorder_fourier},
\begin{equation}\label{eachmode_order2}
\langle \btau^p_{j,2} \rangle  - \frac{\eta_j}{\eta} \langle \gamd_2 \rangle  =
\Re\left\{\frac{\eta_j}{2\eta}\frac{\De_j}{1+i\De_j}
\left( \nabla\widetilde\u_1^{T*}\cdot \widetilde \gamd_1+  \widetilde\gamd_1 \cdot \nabla\widetilde\u_1^* - \widetilde\u_1^* \cdot \nabla \widetilde\gamd_1
\right)
 \right\}\cdot
\end{equation}
We then sum up Eq.~\eqref{eachmode_order2} for each mode, and add the Newtonian contribution from the solvent, to obtain
\begin{equation}
\langle \btau_{2} \rangle  - \langle \gamd_2 \rangle  =
\Re\left\{\left(\sum_{j=1}^N\frac{\eta_j}{2\eta}\frac{\De_j}{1+i\De_j}\right)
\left( \nabla\widetilde\u_1^{T*}\cdot \widetilde \gamd_1+  \widetilde\gamd_1 \cdot \nabla\widetilde\u_1^* - \widetilde\u_1^* \cdot \nabla \widetilde\gamd_1
\right)
 \right\}\cdot
\end{equation}
Using the matrices given in \S\ref{oldroyc_C}, we obtain 
\begin{equation}
\langle \btau_{2} \rangle  - \langle \gamd_2 \rangle  = e^{-2y}
\Re\left\{\left(\sum_{j=1}^N\frac{\eta_j}{\eta}\frac{\De_j}{1+i\De_j}\right)
\left[
\begin{array}{cc}
  6y^2-2y-1   &i (1+2y-2y^2) \\
i(1+2y-2y^2)  &    2y^2+2y+1 
\end{array}
\right]
 \right\},
\end{equation}
and therefore
\begin{equation}\label{order2_matrix}
\langle \btau_2 \rangle - \langle \gamd_2 \rangle=
e^{-2y}\sum_{j=1}^N
\left(\frac{\eta_j}{\eta}\frac{\De_j}{1+\De_j^2}\right)
\left[
\begin{array}{cc}
  6y^2-2y-1   &\De_j(1+2y-2y^2) \\
\De_j(1+2y-2y^2)  &    2y^2+2y+1 
\end{array}
\right]\cdot
\end{equation}
After taking the divergence and curl of Eq.~\eqref{order2_matrix}, we obtain the equation for the mean streamfunction as
\begin{equation}
\nabla^4 \langle \psi_2 \rangle = - \left(\sum_{j=1}^N\frac{\eta_j}{\eta}\frac{\De_j^2}{1+\De_j^2}\right)
\frac{\d ^2 }{d y^2}
\left[e^{-2y}(1+2y-2y^2)\right],
\end{equation}
and calculations similar to those presented in \S\ref{oldroyc_C} lead to the swimming speed of the sheet
\begin{equation}
\frac{U}{U_N} =
1-  \sum_{j=1}^N \frac{\eta_j}{\eta} \frac{\De_j^2}{1+\De_j^2}\cdot
\end{equation}
In the case of a single relaxation time, the result of Eq.~\eqref{finalU} is recovered. Finally, the mechanical efficiency is given by 
\begin{equation}
\frac{{\cal E}}{{\cal E}_N} = 1-  \sum_{j=1}^N \frac{\eta_j}{\eta} \frac{\De_j^2}{1+\De_j^2}\cdot
\end{equation}

\bibliographystyle{unsrt}
\bibliography{SwimmingViscoelastic}
\end{document}